\documentclass[twoside,twocolumn,9pt]{article}
\usepackage{extsizes}
\usepackage[super,sort&compress,comma]{natbib} 
\usepackage[version=3]{mhchem}
\usepackage[left=1.5cm, right=1.5cm, top=1.785cm, bottom=2.0cm]{geometry}
\usepackage{balance}
\usepackage{mathptmx}
\usepackage{sectsty}
\usepackage{graphicx} 
\usepackage{lastpage}
\usepackage[format=plain,justification=justified,singlelinecheck=false,font={stretch=1.125,small,sf},labelfont=bf,labelsep=space]{caption}
\usepackage{float}
\usepackage{fancyhdr}
\usepackage{fnpos}
\usepackage[english]{babel}
\addto{\captionsenglish}{%
  
}
\usepackage{array}
\usepackage{droidsans}
\usepackage{charter}
\usepackage[T1]{fontenc}
\usepackage[usenames,dvipsnames]{xcolor}
\usepackage{setspace}
\usepackage[compact]{titlesec}
\usepackage{hyperref}
\usepackage{tabularx}
\usepackage{booktabs}
\usepackage{amsmath}

\usepackage{epstopdf}

\definecolor{cream}{RGB}{222,217,201}

\begin{document}

\pagestyle{fancy}
\thispagestyle{plain}
\fancypagestyle{plain}{
\renewcommand{\headrulewidth}{0pt}
}

\makeFNbottom
\makeatletter
\renewcommand\LARGE{\@setfontsize\LARGE{15pt}{17}}
\renewcommand\Large{\@setfontsize\Large{12pt}{14}}
\renewcommand\large{\@setfontsize\large{10pt}{12}}
\renewcommand\footnotesize{\@setfontsize\footnotesize{7pt}{10}}
\makeatother

\renewcommand{\thefootnote}{\fnsymbol{footnote}}
\renewcommand\footnoterule{\vspace*{1pt}%
\color{cream}\hrule width 3.5in height 0.4pt \color{black}\vspace*{5pt}} 
\setcounter{secnumdepth}{5}

\makeatletter 
\renewcommand\@biblabel[1]{#1}            
\renewcommand\@makefntext[1]%
{\noindent\makebox[0pt][r]{\@thefnmark\,}#1}
\makeatother 
\renewcommand{\figurename}{\small{Fig.}~}
\sectionfont{\sffamily\Large}
\subsectionfont{\normalsize}
\subsubsectionfont{\bf}
\setstretch{1.125} 
\setlength{\skip\footins}{0.8cm}
\setlength{\footnotesep}{0.25cm}
\setlength{\jot}{10pt}
\titlespacing*{\section}{0pt}{4pt}{4pt}
\titlespacing*{\subsection}{0pt}{15pt}{1pt}

\fancyfoot{}
\fancyfoot[LO,RE]{\vspace{-7.1pt}\includegraphics[height=9pt]{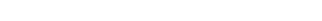}}
\fancyfoot[CO]{\vspace{-7.1pt}\hspace{13.2cm}\includegraphics{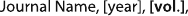}}
\fancyfoot[CE]{\vspace{-7.2pt}\hspace{-14.2cm}\includegraphics{head_foot/RF}}
\fancyfoot[RO]{\footnotesize{\sffamily{1--\pageref{LastPage} ~\textbar  \hspace{2pt}\thepage}}}
\fancyfoot[LE]{\footnotesize{\sffamily{\thepage~\textbar\hspace{3.45cm} 1--\pageref{LastPage}}}}
\fancyhead{}
\renewcommand{\headrulewidth}{0pt} 
\renewcommand{\footrulewidth}{0pt}
\setlength{\arrayrulewidth}{1pt}
\setlength{\columnsep}{6.5mm}
\setlength\bibsep{1pt}

\makeatletter 
\newlength{\figrulesep} 
\setlength{\figrulesep}{0.5\textfloatsep} 

\newcommand{\topfigrule}{\vspace*{-1pt}%
\noindent{\color{cream}\rule[-\figrulesep]{\columnwidth}{1.5pt}} }

\newcommand{\botfigrule}{\vspace*{-2pt}%
\noindent{\color{cream}\rule[\figrulesep]{\columnwidth}{1.5pt}} }

\newcommand{\dblfigrule}{\vspace*{-1pt}%
\noindent{\color{cream}\rule[-\figrulesep]{\textwidth}{1.5pt}} }

\makeatother

\twocolumn[
  \begin{@twocolumnfalse}
{\includegraphics[height=30pt]{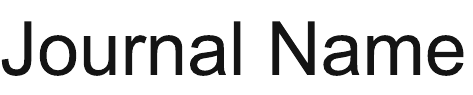}\hfill\raisebox{0pt}[0pt][0pt]{\includegraphics[height=55pt]{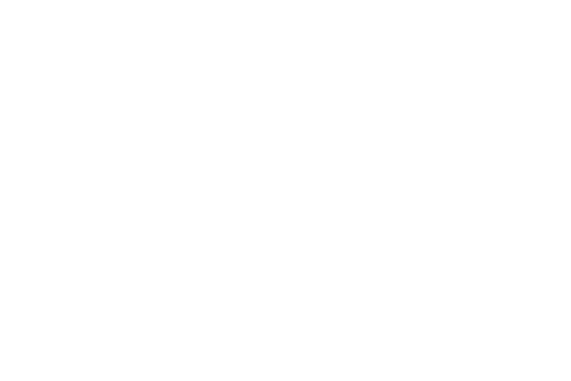}}\\[1ex]
\includegraphics[width=18.5cm]{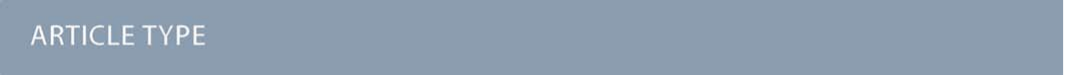}}\par
\vspace{1em}
\sffamily
\begin{tabular}{m{4.5cm} p{13.5cm} }

\includegraphics{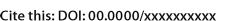} & \noindent\LARGE{\textbf{Safe or Slow? The Illusion of Thermal Stability Under Reduced-Velocity Nail Intrusion}} \\
\vspace{0.3cm} & \vspace{0.3cm} \\

 & \noindent\large{Eymen Ipek,$^{\ddag a}$ Oliver Korak,$^{\ddag b}$ Georg Gsellmann,$^{\ddag b}$ and Andrey Golubkov$^{\ddag \ast b}$} \\

\includegraphics{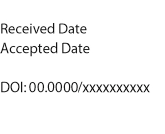} & \noindent\normalsize{This study investigates the effects of nail penetration speed on the safety outcomes of large-format automotive lithium-ion pouch cells. Through six controlled tests varying the speed of nail insertion, we observed that lower penetration speeds did not induce thermal runaway; instead, the cells exhibited self-discharge while the nail remained embedded. These findings suggest that penetration speed is a critical factor in the onset of thermal runaway, providing valuable insights for the development of safer battery systems and more effective safety testing protocols.} \\

\end{tabular}

 \end{@twocolumnfalse} \vspace{0.6cm}

  ]

\renewcommand*\rmdefault{bch}\normalfont\upshape
\rmfamily
\section*{}
\vspace{-1cm}


\footnotetext{\textit{$^{a}$~Graz University of Technology; E-mail: eymen.ipek@student.tugraz.at}}
\footnotetext{\textit{$^{b}$~Virtual Vehicle Research GmbH; E-mail: andrey.golubkov@v2c2.at}}
\footnotetext{\textit{$^{\ddag}$~These authors contributed equally to this work.}}


\section{Introduction}

The increasing prevalence of lithium-ion batteries as the primary energy storage solution in electric vehicles has placed a significant emphasis on ensuring their safety under various operational and accidental conditions. Among the rigorous safety testing protocols employed, mechanical abuse testing plays a crucial role in evaluating the resilience of these high-energy-density devices.

The nail penetration test, a type of mechanical abuse test, is particularly important as it simulates the potential for internal short circuits caused by the intrusion of foreign objects, such as nails or other sharp debris, during vehicle operation or accidents. The high energy density inherent in Li-ion batteries necessitates thorough safety evaluations to protect vehicle occupants and prevent severe consequences arising from battery failure. The nail penetration test, by its nature, represents a severe form of stress that can potentially lead to catastrophic events like thermal runaway if the battery cell's safety mechanisms are insufficient.

Nail speed is considered a crucial parameter in the nail penetration test as it is believed to significantly influence the manner in which the internal short circuit is initiated and the extent of physical damage inflicted upon the battery cell's internal components. The rate at which the nail traverses the cell layers can affect the nature of the contact established between the positive and negative electrodes, as well as the duration of this short-circuiting event. It is plausible that different penetration speeds could result in variations in the area of electrical contact and the temporal profile of the short circuit current, which in turn can impact the subsequent thermal response of the battery.

The phenomenon of thermal runaway takes place when a reaction in the cell exhibits more energy than energy can flux away through diffusion.

In his Master Thesis, Gsellmann investigated the influence of Nail intrusion speed on the TR. The increased time delay between Nail insertion and TR raised the question if there is a limit speed which would not lead to TR.

Conducted experiments in literature and nail penetration standards for automotive battery cells are given in Table~\ref{tab:nail_speed_overview} besides nail penetration parameters used in this study.

\begin{table*}[h!]
\small
\caption{Primary specifications of the cells for testing}
\label{tab:nail_speed_overview}
\begin{tabular*}{\textwidth}{@{\extracolsep{\fill}}ccccc}
\hline
\textbf{Specified Nail Speed} & \textbf{Nail dia.} & \textbf{Casing} & \textbf{Chemistry} & \textbf{Reference} \\
(mm$\cdot$s$^{-1}$) & (mm) & & & \\
\hline
0.001/0.002/0.1/1/10 & 3 & Pouch & NMC & This paper \\
0.02/40 & 1.27/3 & Pouch & LiPo & \cite{huang2020} \\
0.1 & 3 & Pouch & NMC & \cite{gerosa_impact_2025} \\
0.1/10/100 & 6 & Pouch & NMC & \cite{ma2019mechanics} \\
1 & N/A & Cyl. & LMO & \cite{kitoh_100_1999} \\
1 & 2 & Pouch & LCO & \cite{ramadass2014} \\
1 & 2/3/4/5 & Pouch & LCO & \cite{doose2021effects} \\
1/5/10/40/80 & 3 & Pouch & LCO & \cite{diekmann2020} \\
3 & 5 & Pouch & LCO & \cite{yokoshima2018direct} \\
6 & 3/4/8 & Cyl. & NCA & \cite{chen2023} \\
10 & 3 & Pouch & NMC & \cite{abaza2017} \\
10/20/40 & 3/5/8 & Cyl. & LFP & \cite{huang2020thermal} \\
12.3 & N/A & Cyl. & NMC & \cite{saniee2024} \\
$\sim$15 & 4 & Cyl. & NMC & \cite{finegan2017tracking} \\
20/30/40 & 3 & Cyl. & NMC & \cite{mao2018failure} \\
80 & 3 & Pouch & LCO & \cite{xu2019} \\
80 & 3 & Cyl. & NMC/LFP & \cite{reichert2014}\\
80 & 3/4 & Cyl. & NMC & \cite{hildebrand_delayed_2016} \\
80 & 4 & Cyl. & NMC & \cite{braghiroli2023} \\
100 & 3 & Pouch & LMO-NMC & \cite{feng2015} \\
100 & 10 & Pouch & NMC & \cite{abaza2018experimental} \\
200 & 2.5 & Pouch & LCO & \cite{wang2020four} \\
1000 & N/A & N/A & N/A & \cite{dubaniewicz2014} \\
0.1 & 1 & Any & Any & VW PV8450 \cite{chen_review_2021}\\
25 & 5-8 & Pouch & Any & GB/T 31485 \cite{chen_review_2021}\\
80 & 3 & Pouch & Any & SAE J2464 \cite{chen_review_2021}\\
80 & 3 & Pouch & Any & USABCGM \cite{chen_review_2021}\\
80 & 3 & Pouch & Any & SAND2005-3123 \cite{doughty_freedomcar_2006}\\
80 & 3 & Pouch & Any & AIS-048 \cite{abaza2017}\\
\hline
\end{tabular*}
\end{table*}

\subsection{Contribution}
This work explores the influence of nail speed on LIB TR behavior. The investigated speed ranges from known 10 mm$\cdot$s$^{-1}$ to previously undocumented 0.001 mm$\cdot$s$^{-1}$.

\section{Materials and Methods}

Detailed description of the deployed testbed environment \cite{battcave} can be found in Essl et. al.  \cite{essl2020comparing} and Ferdigg and Mair \cite{ferdigg2025nvpf}.

\subsection{Li-Ion Cell}
The experiment used a 66 Ah pouch cell, manufactured by a reputable automotive Li-Ion cell producer \cite{batemoe66a}. It origins from a fresh module, which was dissected in the laboratory. Its primary specifications are given in Table~\ref{tab:cell_specs}. The cell is assumed to feature a graphite-based anode and an NMC cathode.

\begin{table}[h!] 
\small
\caption{Primary specifications of the used cell for testing}
\label{tab:cell_specs}
\begin{tabular*}{0.48\textwidth}{@{\extracolsep{\fill}}ccc}
\hline
\textbf{Parameter}	& \textbf{Value}& \textbf{Unit}\\
\hline
Cathode & NMC712	& -\\
Grav. Energy Density & 259	& Wh/kg\\
Vol. Energy Density & 648	& Wh/l\\
Cell Energy & 240	& Wh\\
Nominal Voltage & 3.66	& V\\
Maximum Voltage & 4.2	& V\\
Cut-off Voltage & 2.5	& V\\
Dimensions & 320x98x11.5	& mm\\
Weight & 890	& g\\
\hline
\end{tabular*}
\end{table}

\subsection{Details of the nail penetration apparatus (NPA) and procedure}

\begin{figure}[h!]
\centering
\includegraphics[width=\linewidth]{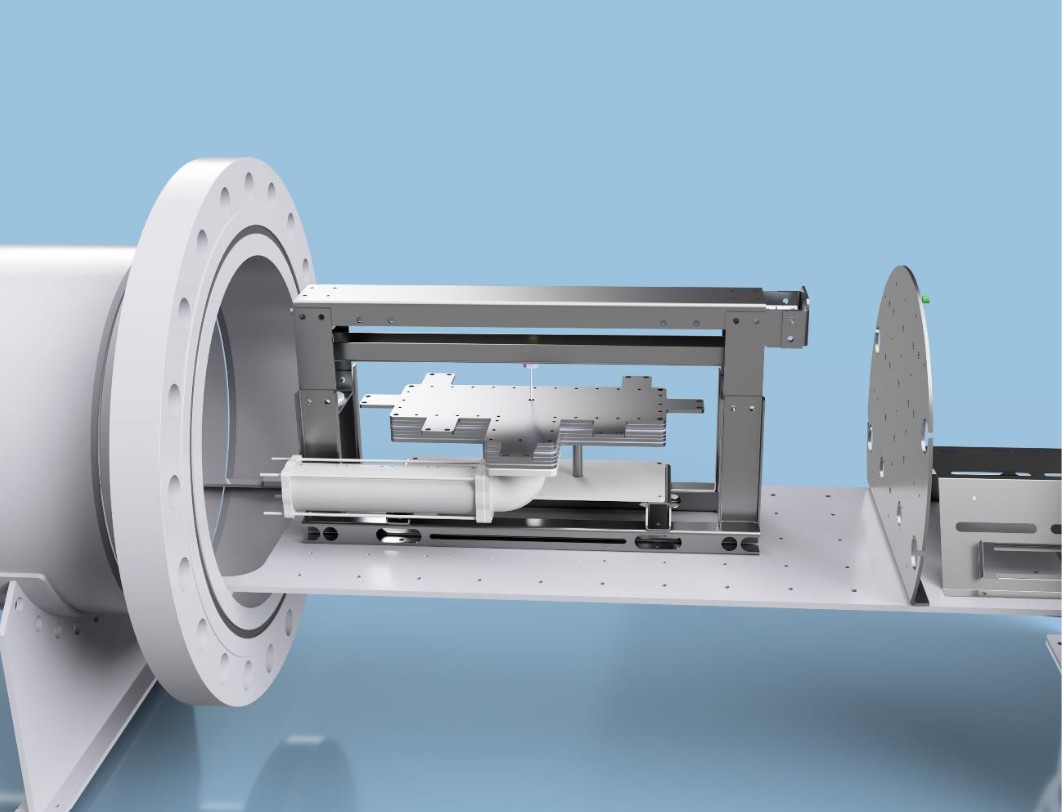}
\caption{Test setup showing the SH mounted on the NPA inside the reactor. The reactor is open in the render picture.\label{Setup_in_Reactor}}
\end{figure}

Experiments were conducted in a closed steel vessel, also termed reactor, under Nitrogen atmosphere. To assess the average temperature of the atmosphere within the vessel, four thermocouples are installed vertically at equal distances inside. Additionally ther vessel is equipped with two gauges. This test series deployed a special sample holder(except the experiment with 0.002mm/s Nailspeed), which allowed to determine the evolved reaction heat through calorimetry. As this is not the focus of this paper and the influence of the sample holder (SH) on the TR behavior is negligible it is not further discussed. The SH consists of a casket where the cell is placed and an exhaust tube equipped with a heat trap. The sample holder is mounted on the NPA, which is placed inside the reactor, as depicted in Fig.~\ref{Setup_in_Reactor}. The NPA allows for a intrusion speed of 50 mm s\textsuperscript{-1} with a force of  4000N on the nail.

\begin{figure}[h!]
\centering
\includegraphics[width=\linewidth]{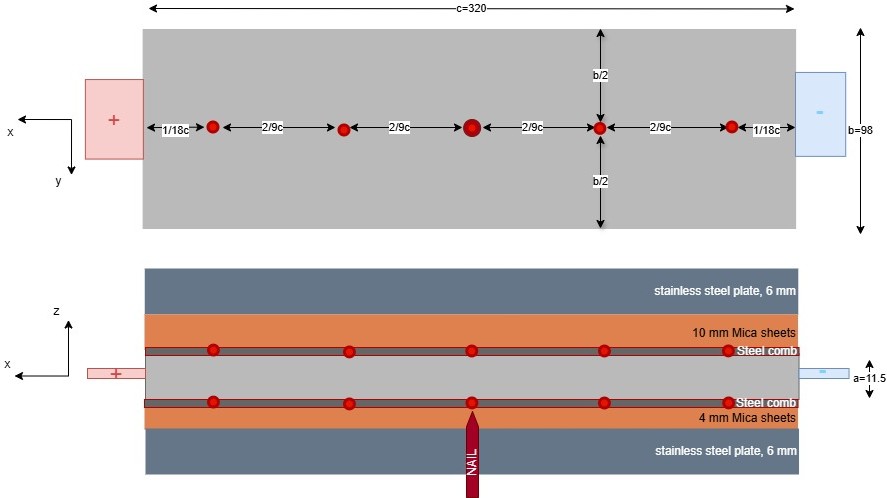}
\caption{TC positions on the cell surface (red dots) and schematic of the stacked setup inside the casket.\label{Cell_setup_Scheme}}
\end{figure}

The cell is housed in a casket where a force of 500 N is exerted on the cell using two springs. Cell expansion is monitored by a displacement transducer. Temperatures on the cell surface are measured with Type K thermocouples. Fig.~\ref{Cell_setup_Scheme} shows the placement of the ten thermocouples on the cell surface.  A pressure sensor monitors pressure built up inside the casket. For thermal isolation, the cell and the TCs are placed between layers of mica sheets. 

\begin{figure}[h!]
\centering
\includegraphics[width=\columnwidth]{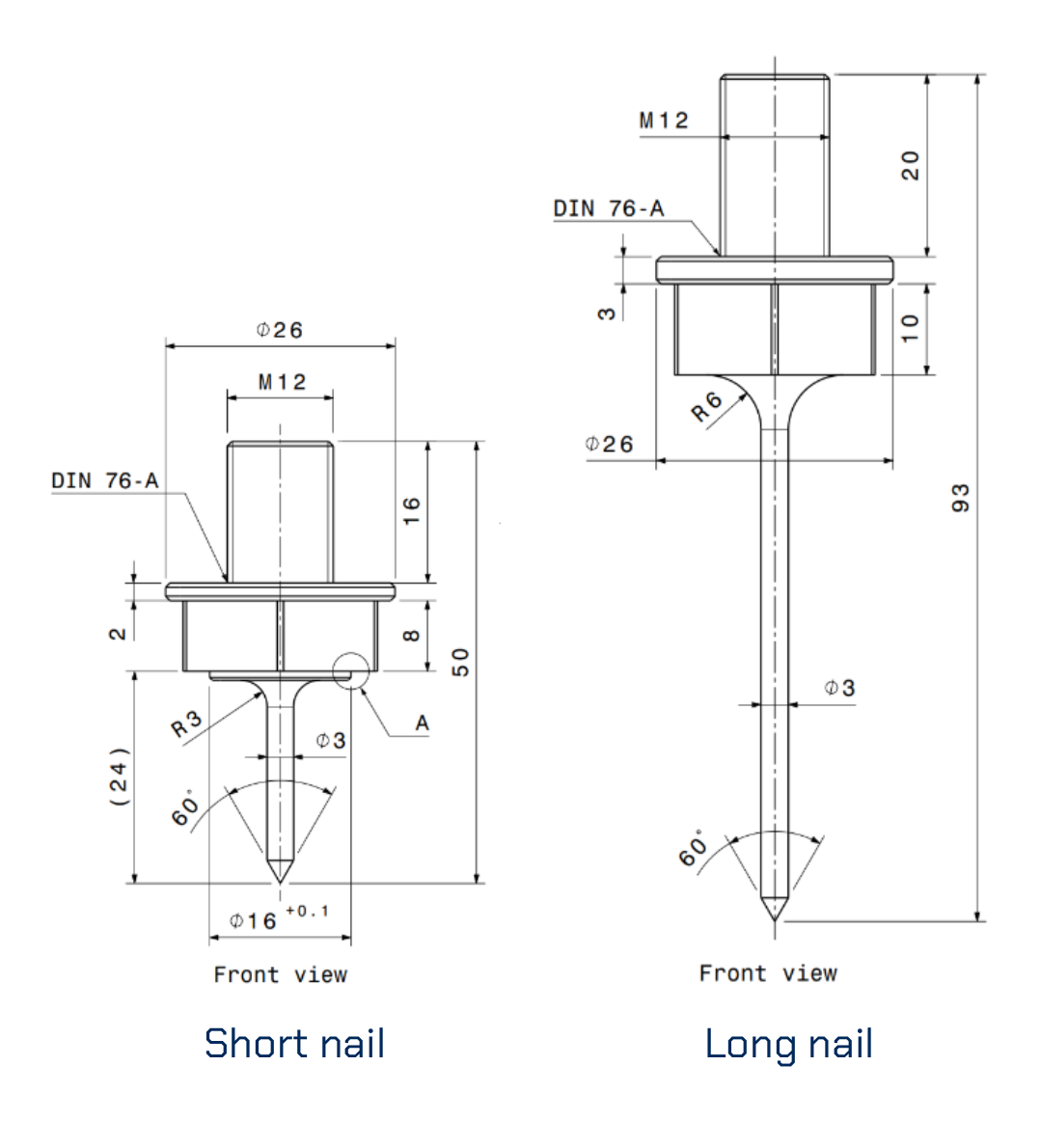}
\caption{Drawings of different nail formats. Long nail was used in the test series.\label{nail_size}}
\end{figure}

Steel grade 42CrMo4 is used as nail material during tests. Fig.~\ref{nail_size} shows examples for a short and a long nail. The long nail format was deployed in this this paper. Nails are screwed into the lever of the NPA and can be easily exchanged. For precise determination of the time of intrusion the potential between the Nail and one of the electrodes is monitored.

The investigated cell is placed inside the casket as depicted in Fig.~\ref{sample_holder}. Cell tabs where elongated with copper sheets to connect charging cables to the cell. Temperature sensor cables and cell tabs are fed through the PTFE frames using silicon foam as sealing material. Fig.~\ref{reactor_before_test} shows the test setup. 

\subsection{Gas Analysis}

A complimentary system of chromatography and infrared spectrometry is deployed to investigate the thermal runaway exhaust gas. The used chromatograph is a µGC Fusion from Inficon. It is equipped with two columns with TCD detectors. The infra red spectrometer was a Bruker Matrix-MG01 with 0.5 cm\textsuperscript{-1} resolution and a 10 cm heated measurement chamber.

\section{Experimental Details}

\begin{figure}[h!]
\centering
\includegraphics[width=0.7\linewidth]{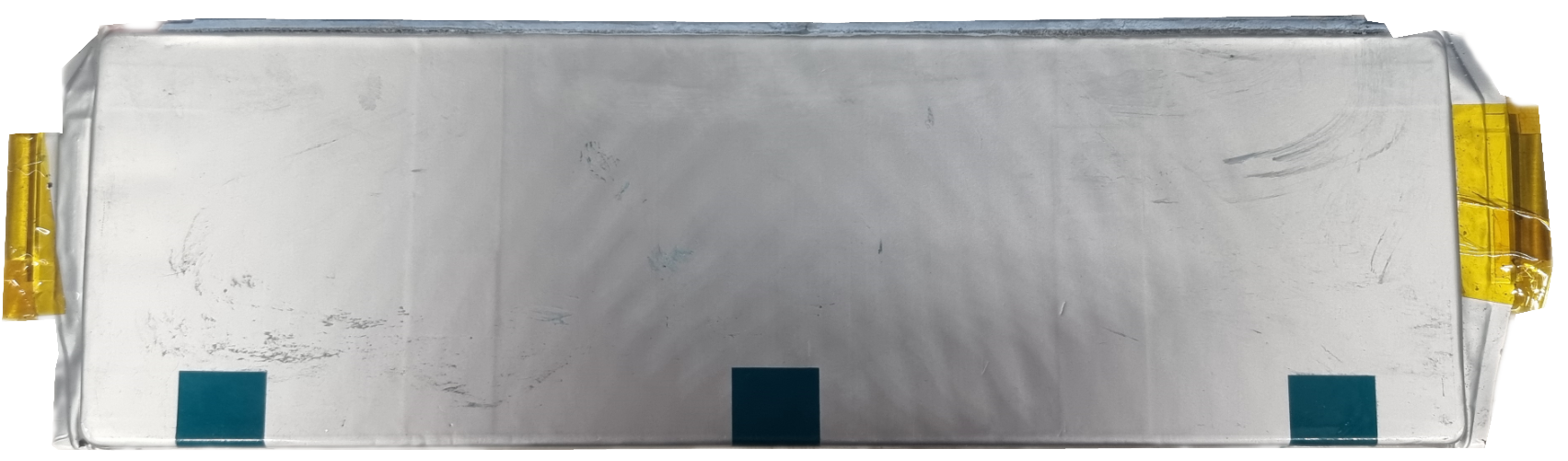}
\caption{Cell before test.\label{cell}}
\end{figure}

Each experiment followed the same experimental procedure. First the cell is checked by measuring the mass and the initial voltage. Pictures are taken from every cell side to scan for visible defects. One of the tested cells is shown in Fig.~\ref{cell}. 

\begin{figure}[h!]
\centering
\includegraphics[width=\linewidth]{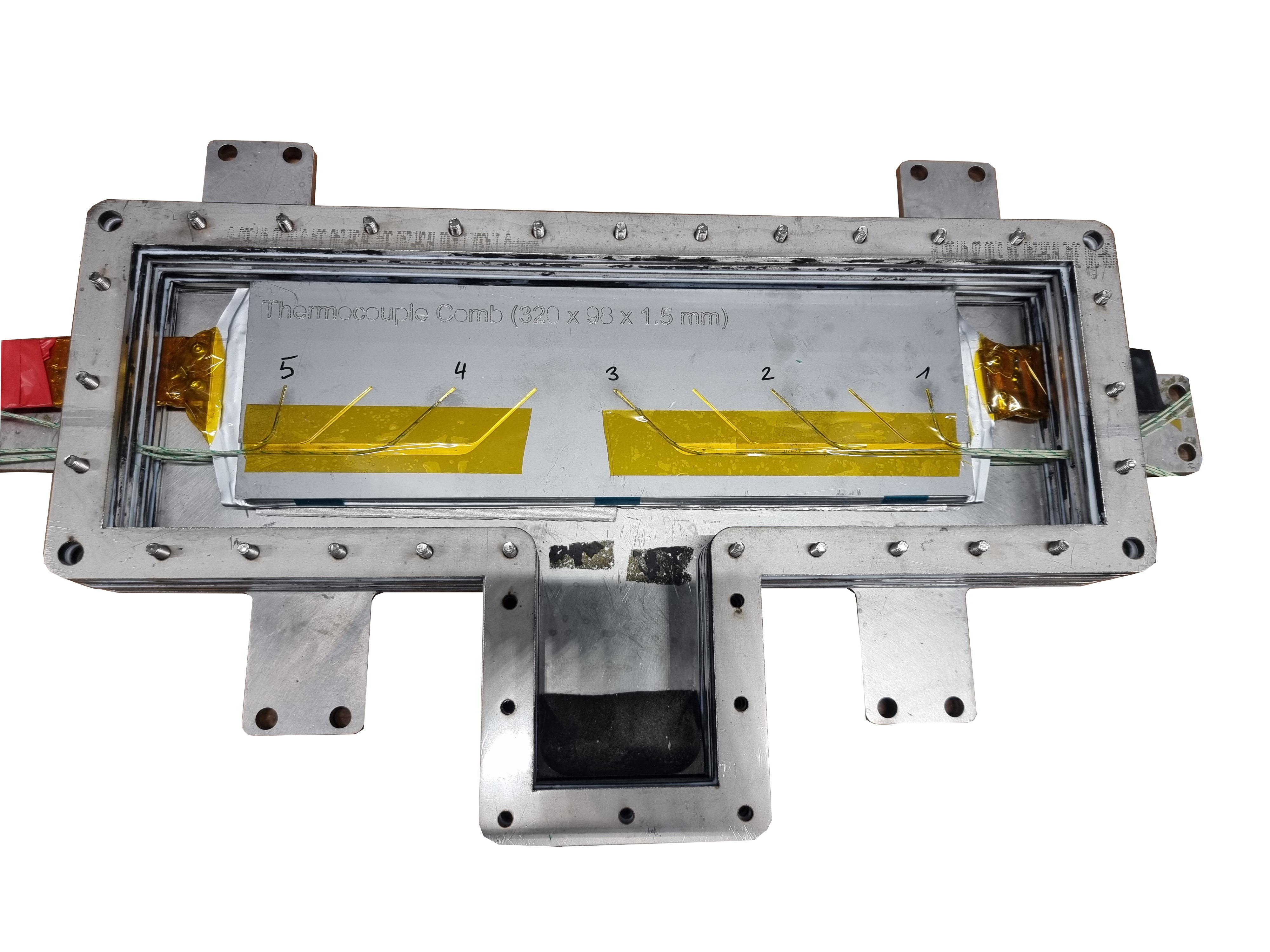}
\caption{Picture of the cell inside the SH. TCs are fixated on the cell by placing them inside a steel comb. Cell terminals are elongated with copper sheets to feed through the stacked SH.\label{sample_holder}}
\end{figure}

Then the cell was placed inside the casket. Copper strips elongated the cell current collectors enabling the connection of cycling cables outside the box. A steel crests held the TCs on the cell surface in place, as depicted in Fig.~\ref{sample_holder}. 

\begin{figure}[h!]
\centering
\includegraphics[width=\linewidth]{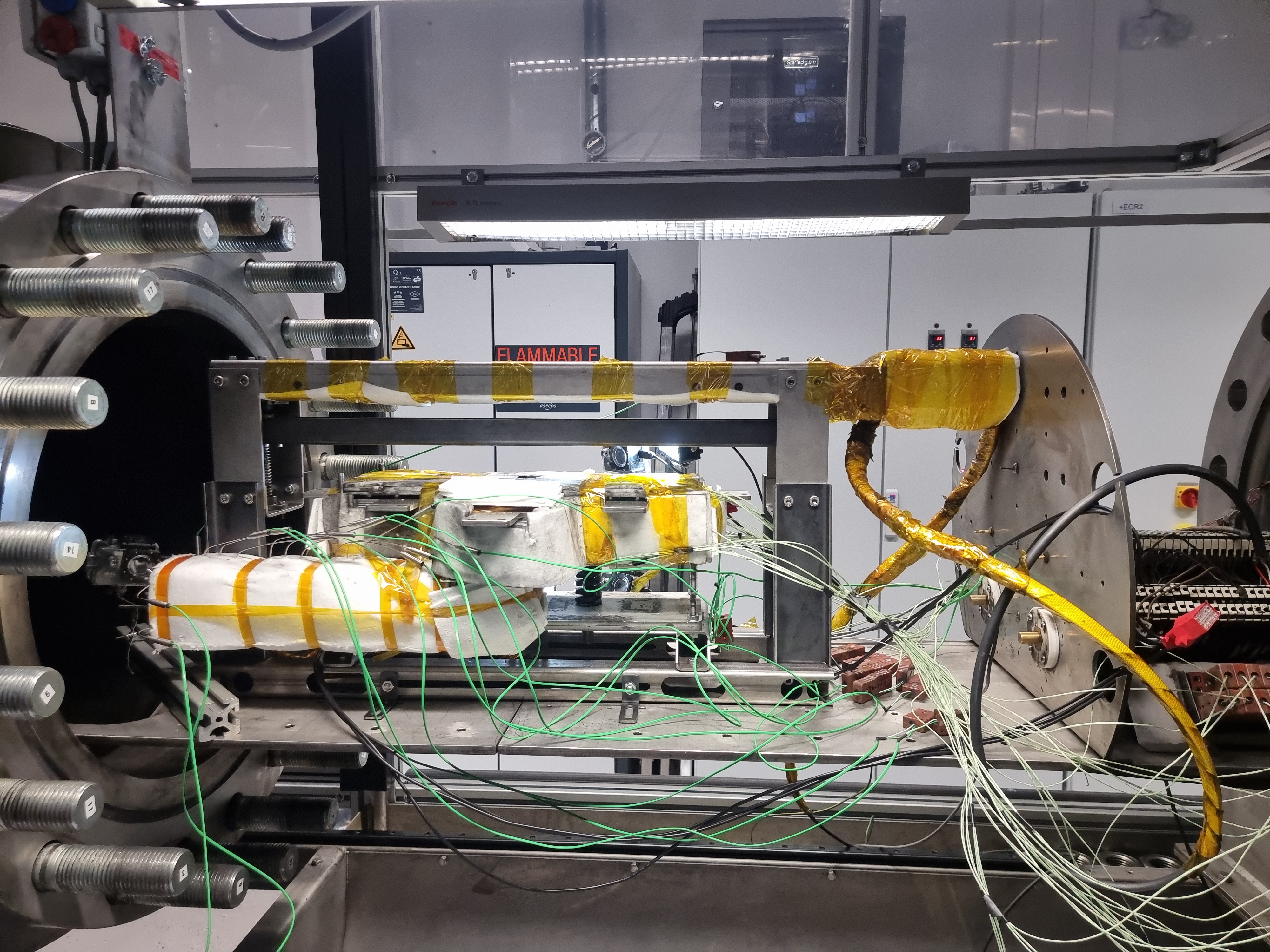}
\caption{Test setup mounted in the reactor before thermal runaway. The calorimeter setup is placed upside down, so the nail penetrates the cell through the bottom plate\label{reactor_before_test}}
\end{figure}

After completion of the stack assembly inside the casket, the casket was closed and a force of 500 N was applied to secure the cell within the sample holder (SH). Additional sensors for pressure and swelling measurements were mounted on the SH prior to wrapping the entire assembly in ceramic insulation foil. The insulation layers were fixed in place using Kapton tape. The assembled setup was then mounted onto the nail penetration apparatus (NPA) within the test bench. 
The cell was subsequently cycled using a constant-current/constant-voltage (CC–CV) protocol. The charge and discharge rate was set to C/3 with a cutoff current of C/20. During the final cycle, the cell was charged to 100\% state of charge (SOC). To record visual data, four action cameras were installed inside the reactor. Fig.~\ref{reactor_before_test} shows the experimental setup prior to closing the reactor vessel for the thermal runaway test. 

To establish an inert atmosphere, the reactor was evacuated twice to 0.01 bar and flushed with nitrogen. An overpressure of 0.2 bar was applied to facilitate leak detection. After measuring the residual oxygen concentration using gas chromatography (GC), the nail trigger was activated. If thermal runaway (TR) was initiated, the system was allowed to rest until a stable temperature condition was reached. Subsequently, the reactor atmosphere was analyzed by directing the gases to the analysis hardware. All gas lines were heated to 170 °C to prevent condensation of analytes. Once the residual oxygen concentration dropped below 2 \%, the NPA was activated. The experiment was monitored via live video feed and real-time data acquisition at 100 Hz.  
TR was determined  by live observation of the cell surface temperature and the pressure of the reactor.

If no thermal runaway was triggered, the cell was either allowed to self-discharge (experiments at 0.001 mm s\textsuperscript{-1}) or actively discharged after a defined observation period (experiments at 0.001 mm s\textsuperscript{-1}). Gas analysis was also performed in the absence of thermal runaway.

\begin{figure}[h!]
\centering
\includegraphics[width=\linewidth]{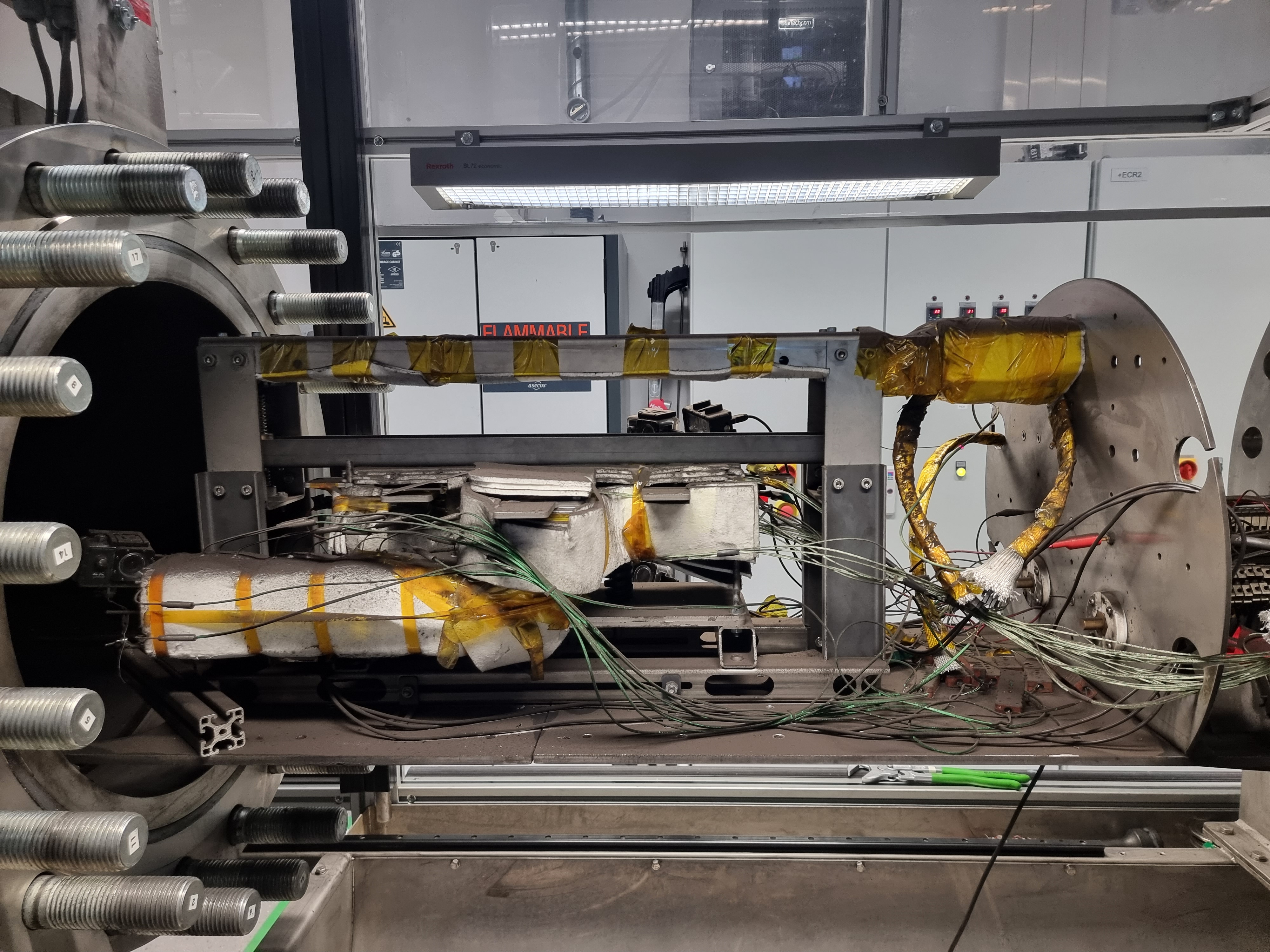}
\caption{Test setup after thermal runaway.\label{reactor_after_test}}
\end{figure}

Before opening the reactor, reaction gases were removed by evacuating the vessel and flushing it with compressed air until no reaction products were detected by gas analysis. Fig.~\ref{reactor_after_test} shows the reactor after a thermal runaway experiment. Prior to closing the reactor again, all electronic components were removed, and the reactor was baked out under vacuum to eliminate residual harmful volatile compounds.

\subsubsection{Parameters varied during testing}
The primary parameter varied during the experimental campaign was the nail penetration speed. Penetration velocities ranged from ultra-low speeds of $0.001~\mathrm{mm\,s^{-1}}$ up to $10~\mathrm{mm\,s^{-1}}$. All other parameters, including nail geometry, cell type, state of charge, ambient atmosphere, and test setup, were kept constant.

\section{Results}

\begin{figure}[h!]
\centering
\includegraphics[width=\linewidth]{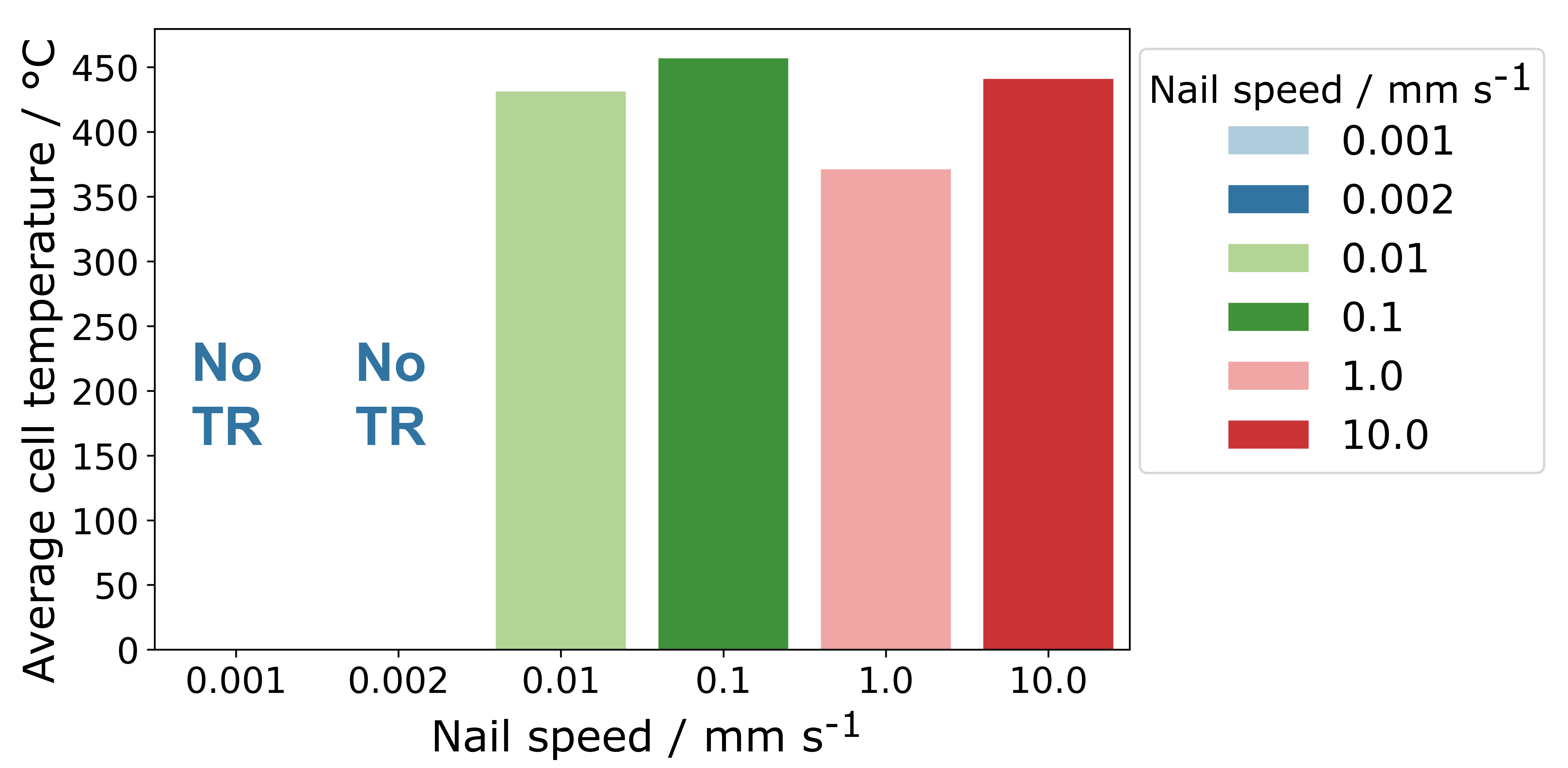}
\caption{Average cell temperature during tests.\label{avg_temp}}
\end{figure}

In case TR was achieved, the average cell temperature reached x +/- X °C. As depicted in Fig.~\ref{avg_temp} we discovered no connection between the Nail speed and the average cell surface temperature during TR. This is comparable to literature results.

\begin{figure}[h!]
\centering
\includegraphics[width=\linewidth]{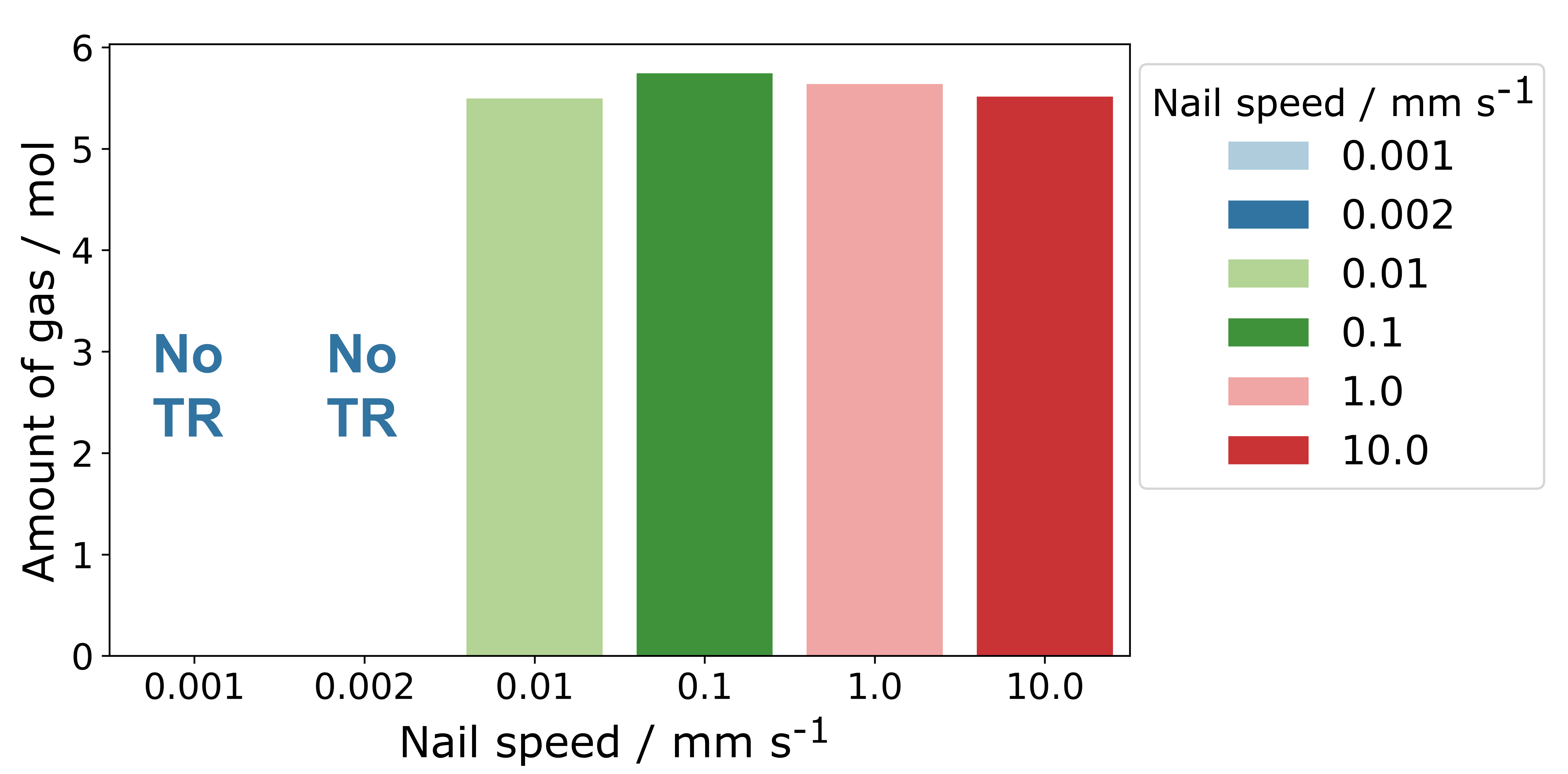}
\caption{Amount of gas in mol.\label{amount_gas_release}}
\end{figure}

In case TR was achieved, an average of x +/- x mol of gas was released from the cell, equal to x +/- x mol/Ah . As Fig.~\ref{amount_gas_release} shows, there was no influence of Nail speed on the amount of released gas observed.

\begin{figure}[h!]
\centering
\includegraphics[width=\linewidth]{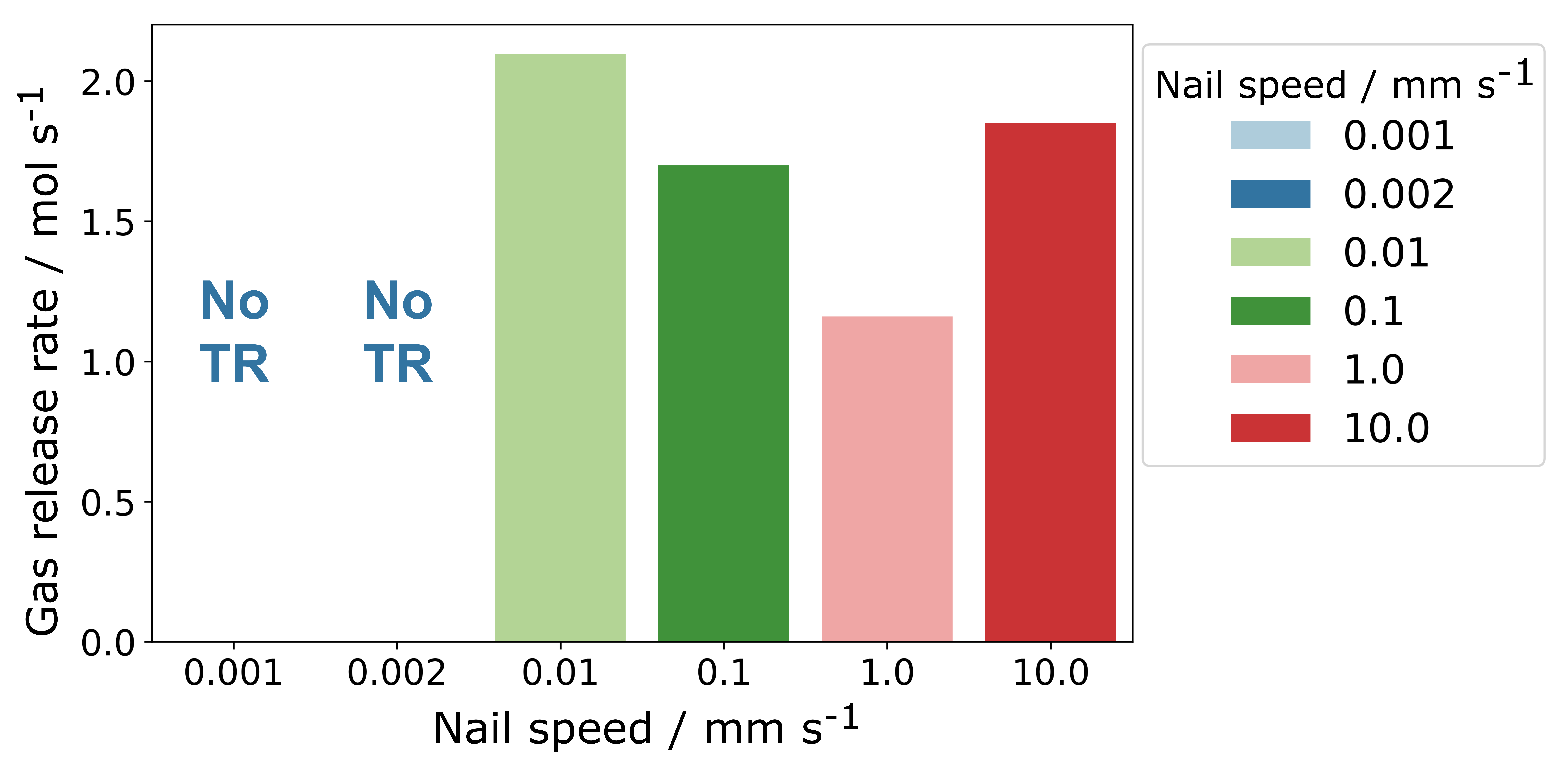}
\caption{Gas release rate during tests.\label{gas_release_rate}}
\end{figure}

In case TR was triggered a difference in the gas release rate was observed, although the results were inconclusive. As Fig.~\ref{gas_release_rate} shows the gas release rate decreases from 0.01 to 1 mm s\textsuperscript{-1} from 2 mol s\textsuperscript{-1} to 1  mol s\textsuperscript{-1} but increase at a Nail speed of 10mm s\textsuperscript{-1} to 1.5 mol s\textsuperscript{-1}. More tests are needed for a unequivocal answer.

The maximum average temperature \( T_{\text{max}} \) is calculated as the maximum value of the average temperature across all \( N \) temperature sensors at a given time \( t \). The formula is given by:
\begin{equation}
T_{\text{max}} = \max\left(\frac{1}{N} \sum_{i=1}^{N} T_{\text{cell}_i}(t)\right)
\end{equation}

\noindent
where:
\( T_{\text{max}} \) is the maximum average temperature, \( N \) is the number of temperature sensors, \( T_{\text{cell}_i}(t) \) is the temperature of the \( i \)-th sensor at time \( t \), and \( t \) is the time.

\begin{figure}[h!]
\centering
\includegraphics[width=\linewidth]{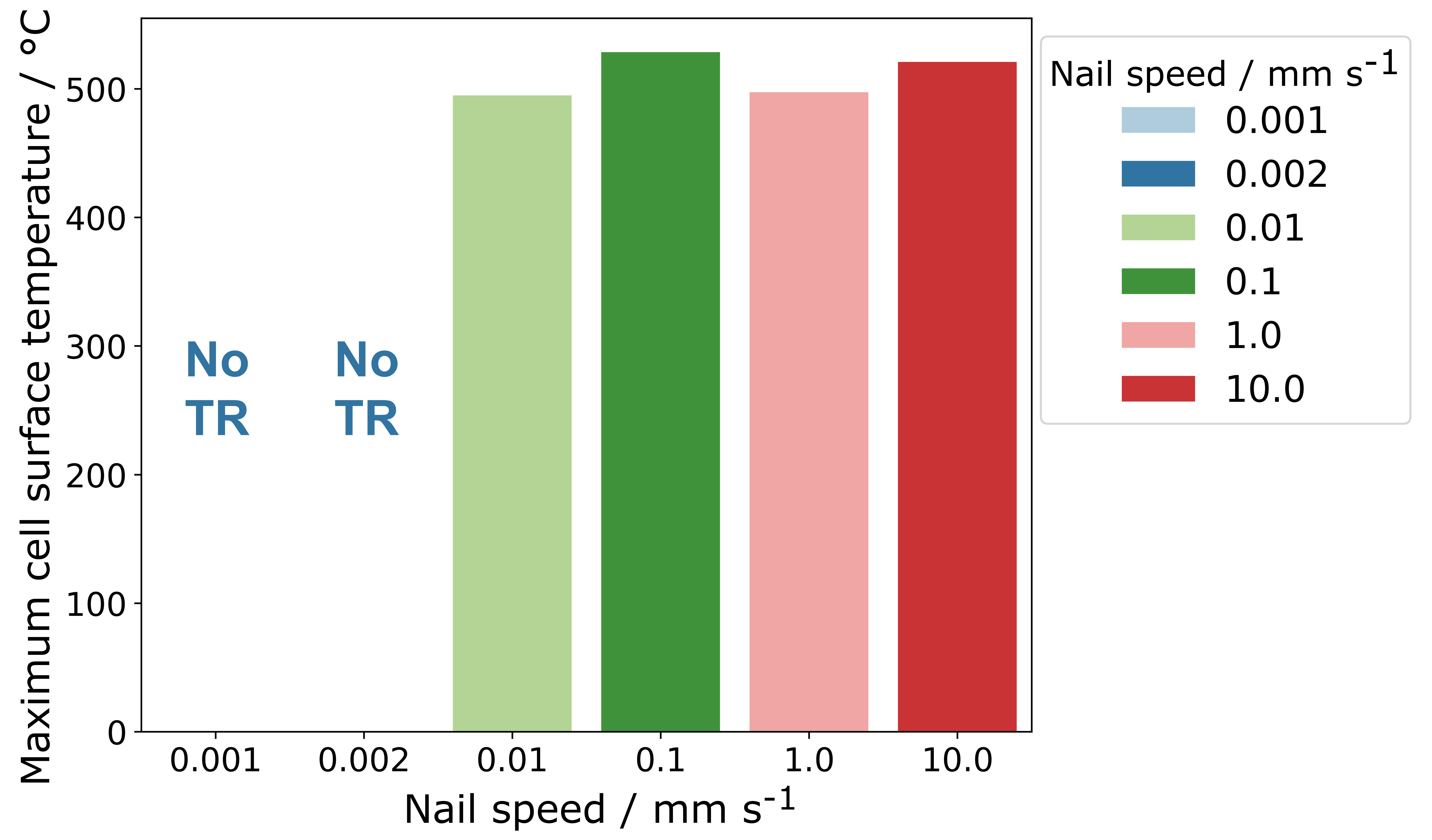}
\caption{\( T_{\text{max}} \), maximum cell surface temperatures according to Equation (1).\label{cell_surface}}
\end{figure}

\begin{figure}[h!]
\centering
\includegraphics[width=\linewidth]{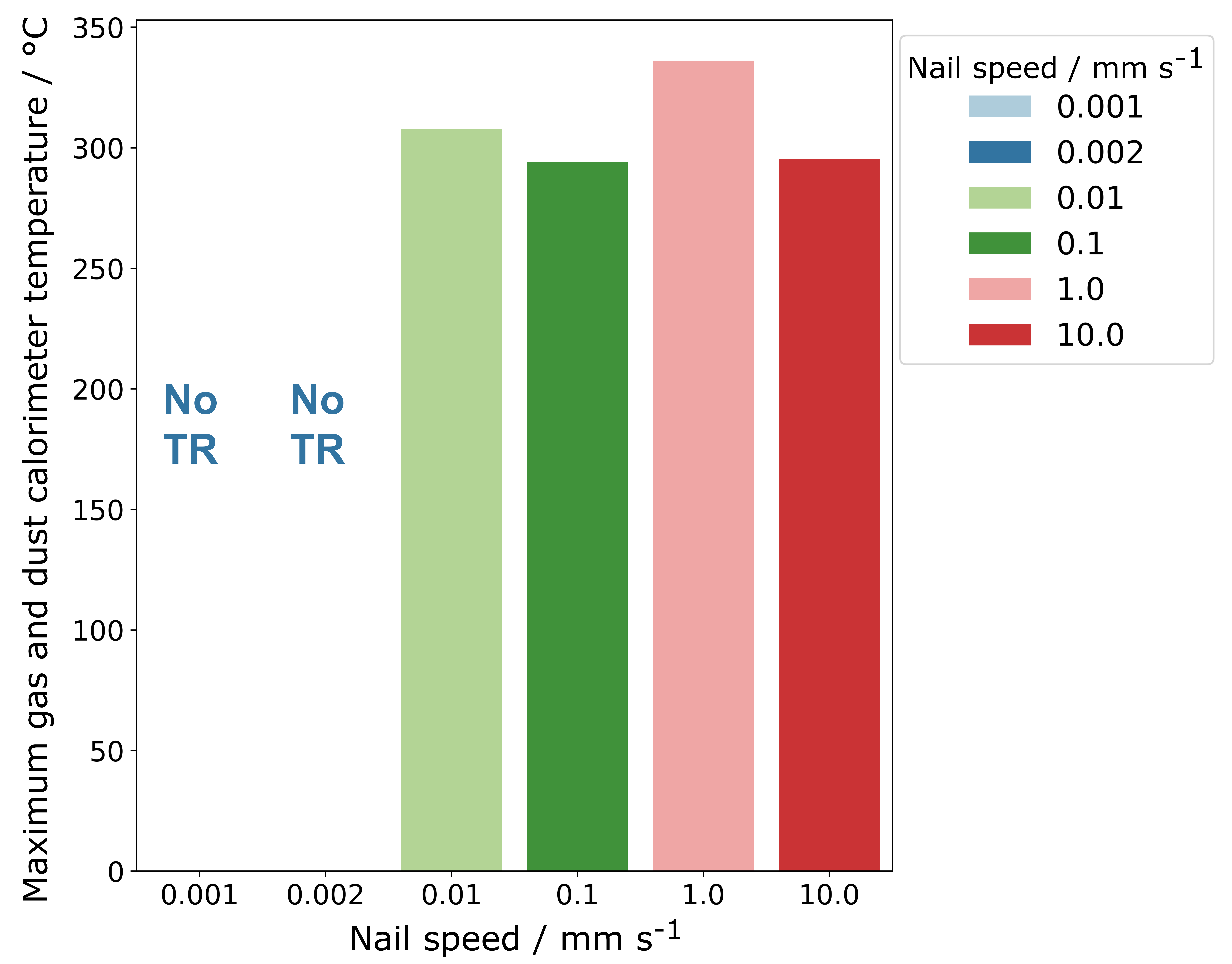}
\caption{\( T_{\text{max}} \), maximum gas and dust calorimeter temperatures according to Equation (1).\label{gas_dust_cal}}
\end{figure}

\begin{figure}[h!]
\centering
\includegraphics[width=\linewidth]{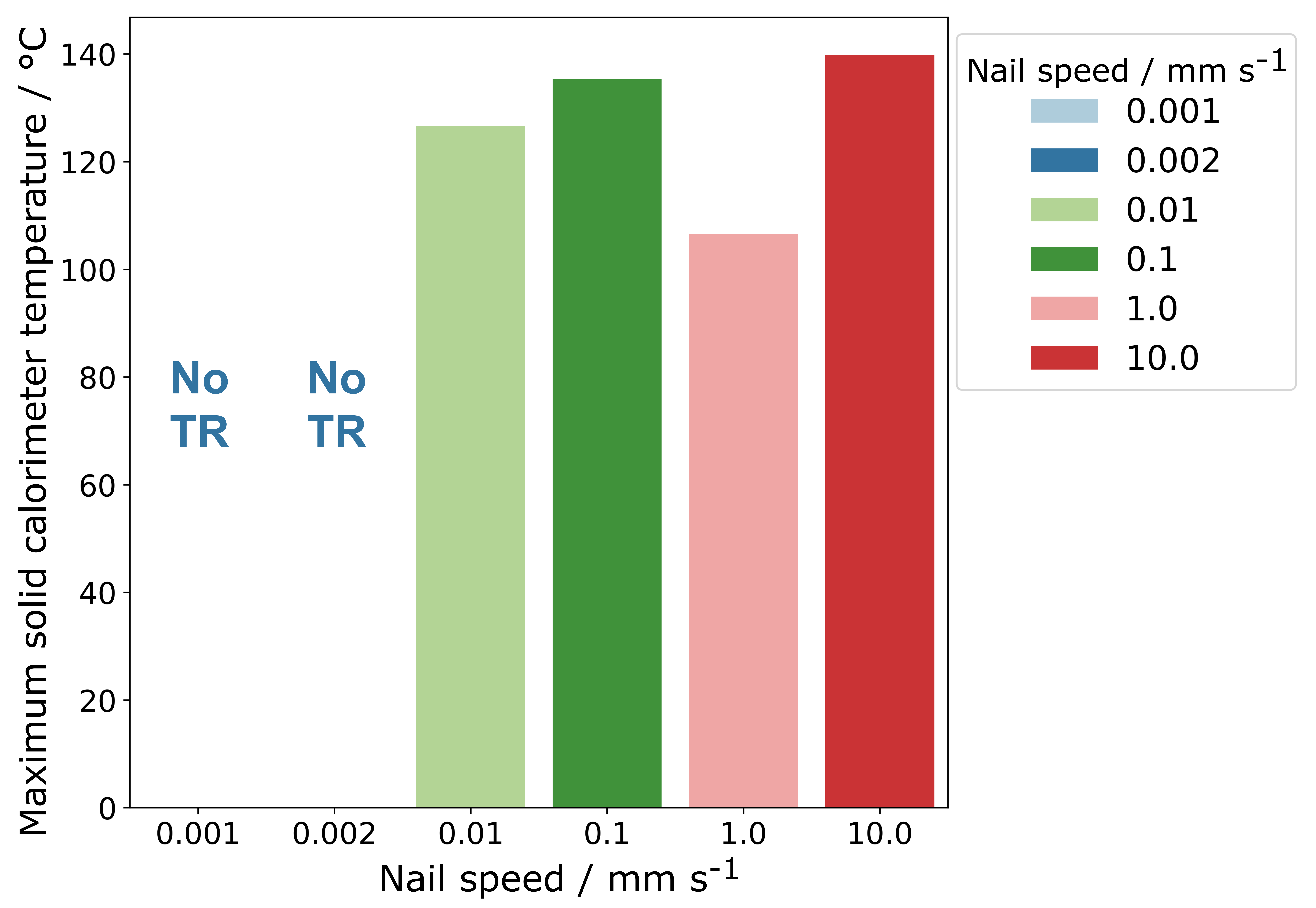}
\caption{\( T_{\text{max}} \), maximum solid calorimeter temperatures according to Equation (1).\label{solid_cal}}
\end{figure}

Fig.~\ref{cell_surface}, Fig.~\ref{gas_dust_cal} and Fig.~\ref{solid_cal} represent maximum average temperatures of cell surface, gas and dust calorimeter and solid calorimeter respectively in the tests where thermal runaway happened according to Equation (1). It must be noted that the solid calorimeter was not calibrated. 

The slight increase in the gas and dust calorimeter at 1 mm s\textsuperscript{-1} could be connected with the slower gas release rate during this experiment. 

\begin{table}[h!] 
\small
\caption{Safety relevant parameters of all tests that experiences a thermal runaway.\label{tab2}}
\begin{tabular*}{\columnwidth}{@{\extracolsep{\fill}}cccc}
\hline
\textbf{Safety relevant parameters}	& \textbf{Mean}	& \textbf{Median}\\
\hline
Average cell temperature / °C & 425$\pm$37	& 436 \\
Average cell surface temperature / °C & 510$\pm$17	& 509\\
Average gas and dust cal. temp. / °C & 308$\pm$20	& 301  \\
Average solid calorimeter temp. / °C & 127$\pm$15	& 131  \\
Gas release / mol		 & 5.60$\pm$0.12	& 5.58 \\
Gas release rate / $\text{mol s}^{-1}$ & 1.70$\pm$0.40	& 1.77 \\
\hline
\end{tabular*}
\end{table}

\begin{figure}[h!]
\centering
\includegraphics[width=\linewidth]{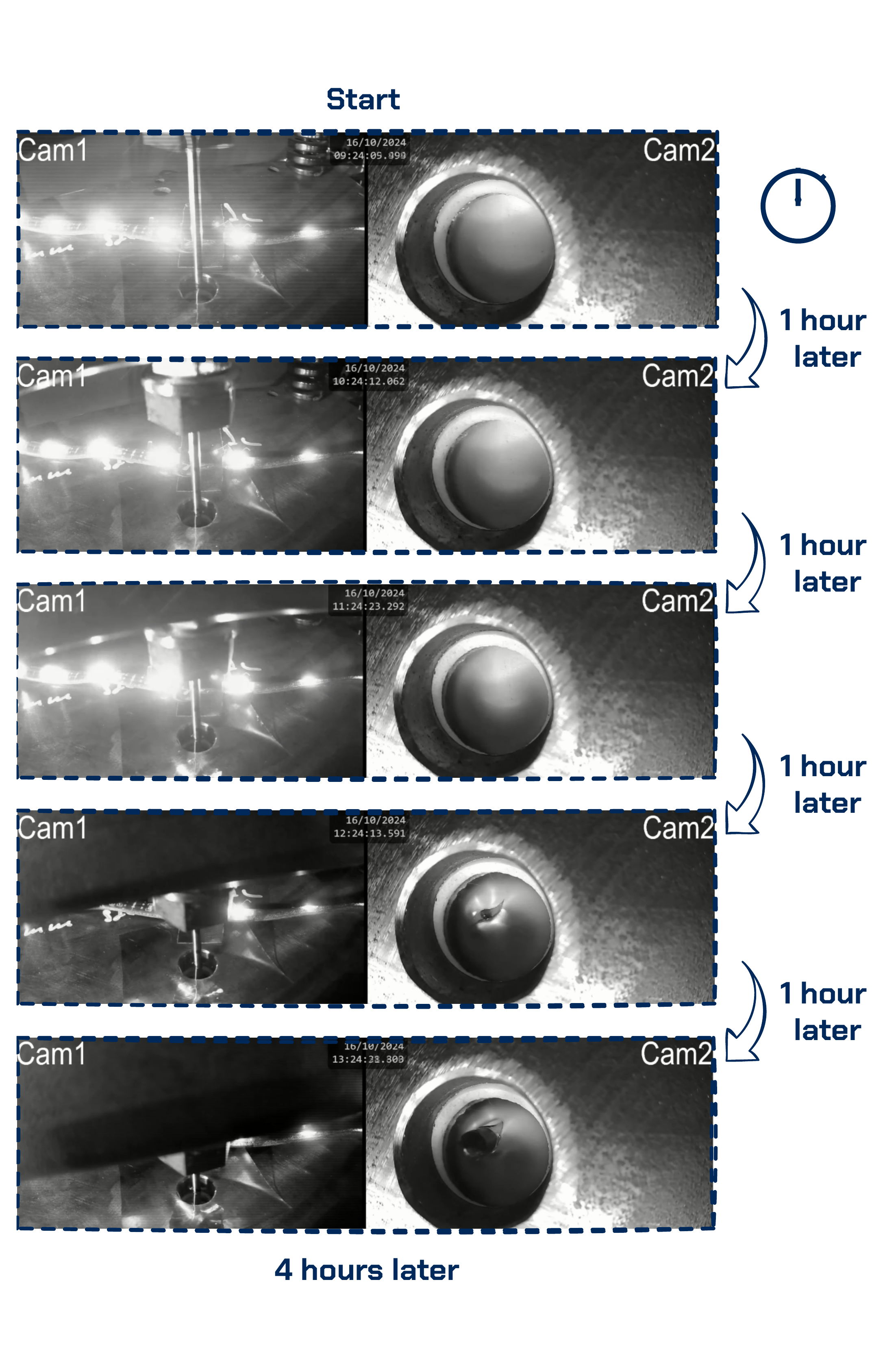}
\caption{The captions during nail penetration tests where the nail speed is 0.002 mm/s.\label{video_snap}}
\end{figure}

Fig.~\ref{video_snap} shows the camera shots during the test where the nail speed was 0.002 mm/s.

\begin{figure}[h!]
\centering
\includegraphics[width=\linewidth]{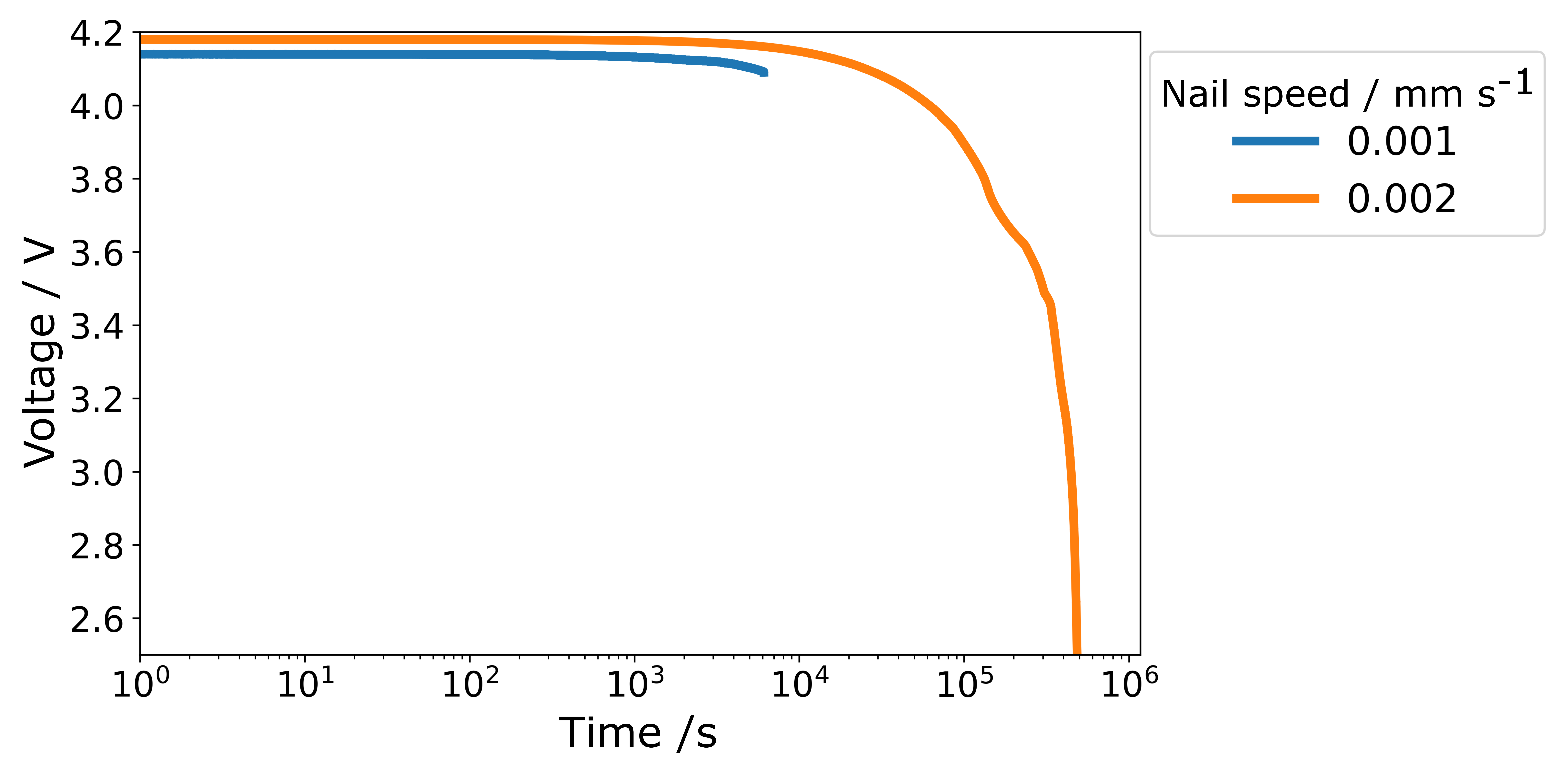}
\caption{Voltage drop due to internal short circuit during the tests where no thermal runaway happened.\label{volt_drop}}
\end{figure}

In the tests where the nail speed were 0.001 and 0.002 mm/s, there was no thermal runaway. However, cell discharged internally. The cell terminals were not connected to the earth. Therefore, the estimated self discharge happened only between inner layer of the cell. Fig.~\ref{volt_drop} shows the self discharge of the cells while estimated parameters are shown in Table~\ref{tab3}.

\begin{table}[h!] 
\small
\caption{Estimated self discharge parameters considering discharge time and discharged energy according to SOC-OCV table of the cell.\label{tab3}}
\begin{tabular*}{0.48\textwidth}{@{\extracolsep{\fill}}ccc}
\hline
\textbf{Nail speed}	& \textbf{Est. self discharge}	& \textbf{Est. self discharge}\\
($\text{mm s}^{-1}$) & \textbf{resistance} (Ohm) & \textbf{current} (A) \\
\hline
0.001 & 1.33	& 2.75\\
0.002 & 7.46	& 0.49\\
\hline
\end{tabular*}
\end{table}

\begin{figure}[h!]
\centering
\includegraphics[width=\linewidth]{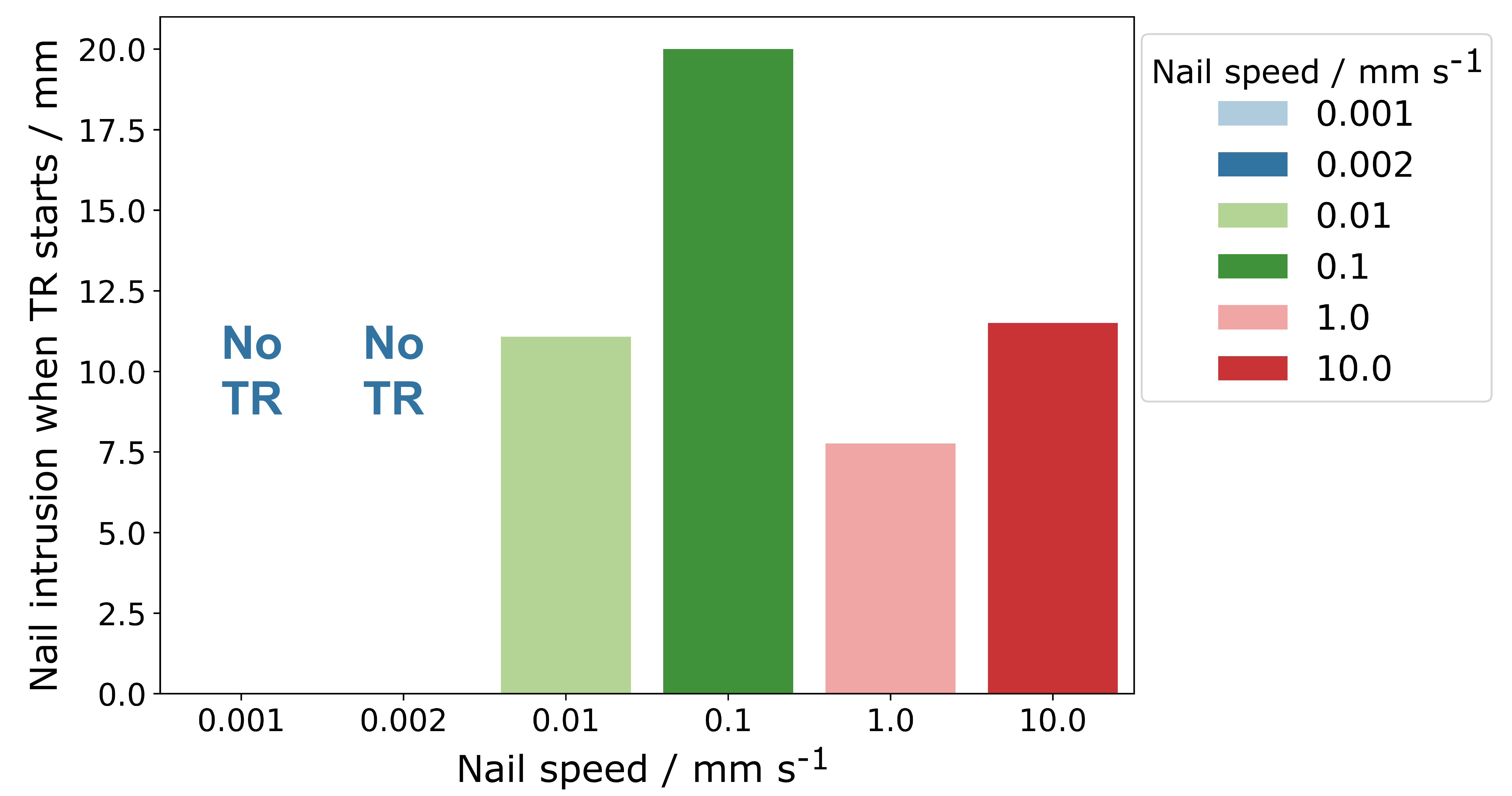}
\caption{Nail intrusion in mm when thermal runaway starts.\label{nail_intr}}
\end{figure}

Fig.~\ref{nail_intr} shows the nail intrusion depth at the onset of thermal runaway as a function of nail penetration speed.

\begin{figure}[h!]
\centering
\includegraphics[width=\linewidth]{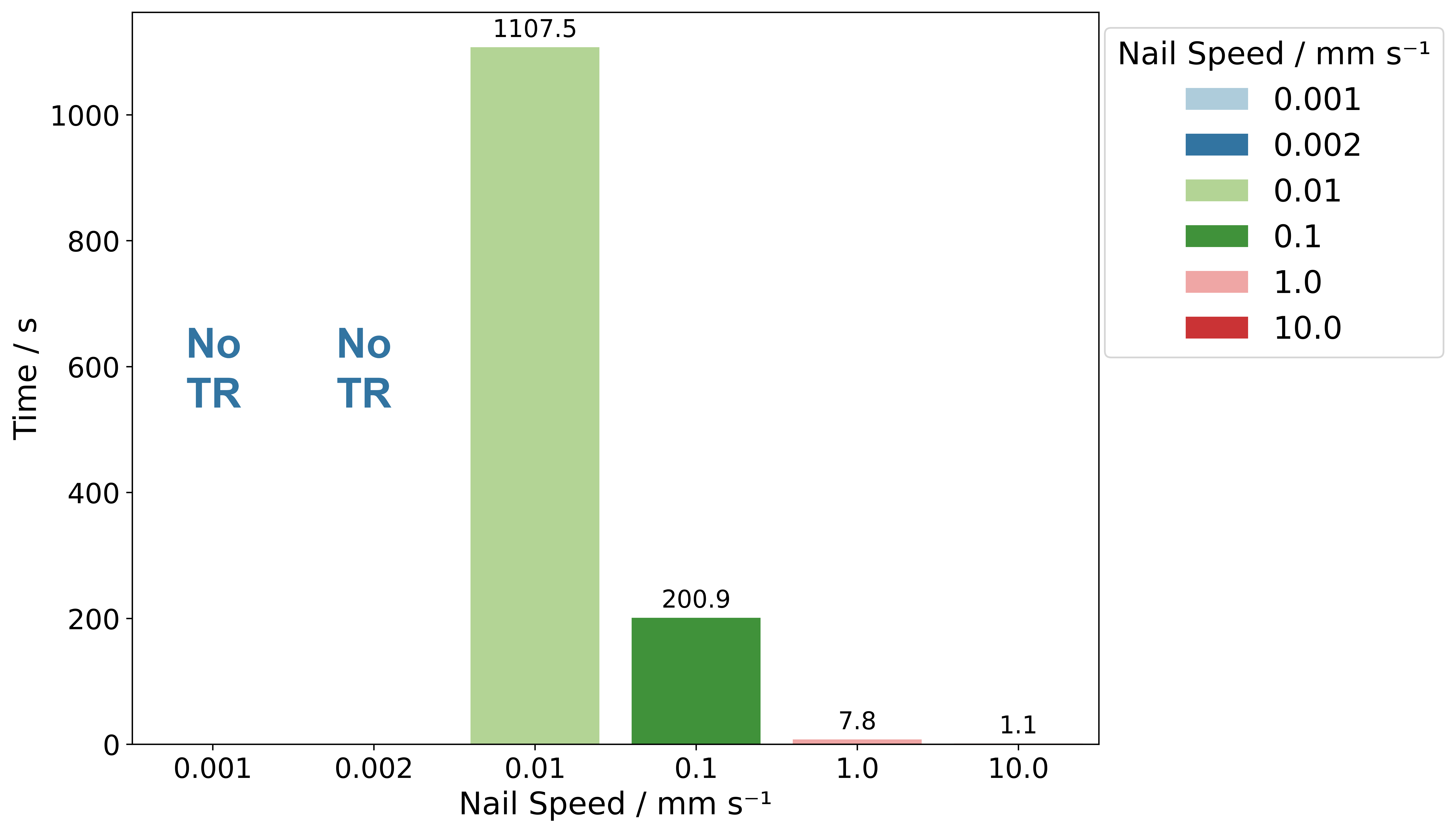}
\caption{Time until nail intrusion starts thermal runaway.\label{delay_nail_entrance}}
\end{figure}

Fig.~\ref{delay_nail_entrance} shows the time elapsed from the start of nail penetration until the initiation of thermal runaway for different nail speeds.

\begin{figure}[h!]
\centering
\includegraphics[width=\linewidth]{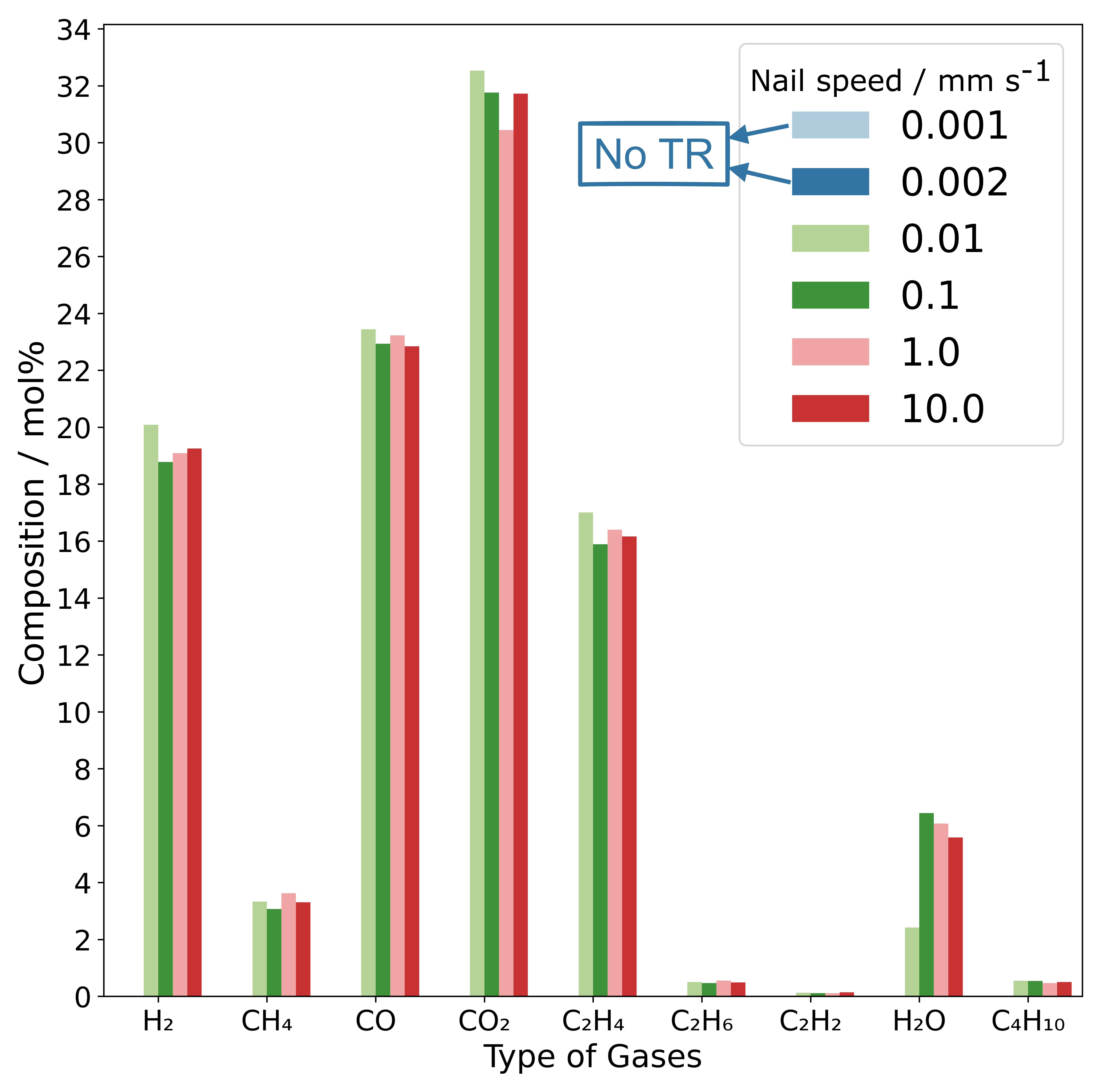}
\caption{Gas composition of vent gas released by cell during the tests.\label{gas_analysis}}
\end{figure}

Fig.~\ref{gas_analysis} shows the composition of vent gas released during the tests for different nail penetration speeds.

The experiments revealed a strong dependence of thermal runaway behavior on nail penetration speed. At very low intrusion speeds ($0.001$ and $0.002~\mathrm{mm\,s^{-1}}$), no thermal runaway was observed. Instead, the cells exhibited internal self-discharge while the nail remained embedded. For penetration speeds of $0.01~\mathrm{mm\,s^{-1}}$ and above, thermal runaway consistently occurred. In cases where thermal runaway was triggered, maximum temperatures, gas release amounts, and gas release rates were comparable across different nail speeds.

\section{Discussion}
The results demonstrate that slow nail penetration can create internal short circuits that remain non-catastrophic, leading primarily to gradual energy dissipation rather than thermal runaway. This behavior creates the illusion of improved safety, although it is strongly dependent on the applied intrusion rate. Once thermal runaway is initiated, the severity of the event appears largely independent of nail speed. These findings highlight the importance of clearly defining test parameters when comparing nail penetration results across studies and standards.

\section{Conclusions}
This study shows that nail penetration speed is a critical parameter influencing whether thermal runaway occurs in lithium-ion pouch cells. Extremely low penetration speeds can prevent thermal runaway by enabling controlled internal discharge, while higher speeds consistently trigger catastrophic failure. However, when thermal runaway does occur, its thermal and gaseous characteristics are largely unaffected by nail speed. These insights underline the need for standardized test definitions and caution against interpreting slow-rate nail penetration as inherently safer without considering real-world abuse scenarios.

\section*{Author contributions}
G.G., and A.G. conceived, designed and performed the experiments; E.I. and O.K. analyzed the data; E.I. and O.K. wrote the paper, A.G. reviewed.

\section*{Conflicts of interest}
There are no conflicts to declare.

\section*{Acknowledgements}
This work originates from the research project SafeLIB. The COMET Project SafeLIB is funded within the framework of COMET—Competence Centers for Excellent Technologies (Grant agreement No. 882506) by BMK, BMDW, the Province of Upper Austria, the Province of Styria, as well as SFG. The COMET Program is managed by FFG.

\footnotesize{

}

\end{document}